# State Independent Proof of Kochen-Specker Theorem with Thirty Rank-Two Projectors

S.P.Toh*

*Faculty of Engineering, The University of Nottingham Malaysia Campus,
Jalan Broga, 43500 Semenyih, Selangor Darul Ehsan, Malaysia.*

(Dated: June 7, 2018)

The Kochen-Specker theorem states that noncontextual hidden variable theories are incompatible with quantum mechanics. We provide a state independent proof of the Kochen-Specker theorem using the smallest number of projectors, i.e., thirty rank-2 projectors, associated with the Mermin pentagram for a three-qubit system.

Contextuality is one of the classically unattainable features of quantum mechanics (QM). The results of measurements in QM depend on context and do not reveal pre-existing values. A context is a set of maximally collection of compatible observables. The contextual QM thus means that the results of measurements in QM depend on the choice of other compatible measurements that are carried out previously or simultaneously. In contrast, classical physics demands that the properties of a system have pre-determined values which are independent of the measurement context. The contradiction between contextual QM and noncontextual classical physics is expressed via Kochen-Specker (KS) theorem. More specifically, the KS theorem states that the predictions of QM are in conflict with the noncontextual hidden-variable (NCHV) theories. The simplest system that can be used to prove KS theorem is a single qutrit. As a qutrit does not refer to nonlocality, it shows that KS theorem is a more general theorem compared to the Bell theorem that rules out the local hidden variable model of QM.

The possibility of testing KS theorem experimentally was once doubted due to the finiteness in measurement times and precision [1, 2]. Cabello [3] and others [4] suggested how KS theorem might be experimentally tested by deriving a set of noncontextual inequalities that are violated by QM for any quantum states but are satisfied by any NCHV theories. Recently, there are many successful experiments that show the violation of noncontextual inequality, for example the experiments on a pair of trapped ions [5], neutrons [6], single photons [7], two photonic qubits [8] and nuclear spins [9]. Apart from the quantum sources, thermal light source has also been used to perform the KS experiment [10].

For the theoretical proofs of the KS theorem, the original version involves 117 directions in three-dimensional real Hilbert space [11]. Peres [12] found simpler proofs with 33 and 24 rays for three- and four-dimensional systems, respectively. Mermin [13] used an array of nine observables for two spin-$\frac{1}{2}$ particles to show quantum contextuality. Similar mathematical simplicity is also shown in KS theorem proof for the three-qubit eight-dimensional system using ten observables [13]. Up to now the smallest numbers of rays required in the proof of KS theorem are 31 [14], 18 [15] and 36 [16] for three-, four- and eight-dimensional systems, respectively.

In this work we provide a state independent proof of the KS theorem in three-qubit system using only 30 rank-2 projection operators (projectors). Rank-1 and rank-2 projectors are also called rays and planes, respectively. The rank-2 projectors used in our proof are generated from the aforementioned ten operators in three-qubit system [13]. To understand how we conceived our proof, it is useful to review, as what we start to do in the next paragraph, some of the KS theorem proofs proposed previously that based on same kind of system.

The above-mentioned ten operators in three-qubit system form five measurement contexts and all the operators in the same context are mutually commute. The Mermin pentagram in Figure 1 depicts clearly the relationships between measurement contexts. Be aware about the shorthand used for denoting operators in Figure 1, for example, $\sigma_x^1$ means the tensor product of $\sigma_x \otimes I \otimes I$. There are two important features of the Mermin pentagram, i.e., (1) the product of the operators in a context gives $I$, an eight-dimensional identity matrix, except the one that is placed horizontally which gives $-I$, (2) each of the operators occurs in two different contexts.

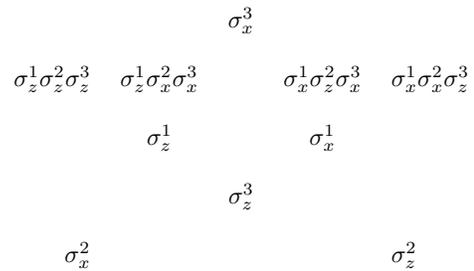

FIG. 1: Mermin pentagram

It was known that QM obeys product rule, which means that for mutually commuting operators $L$, $M$ and $N$ that related via $LM = N$, the value functions, $v(\cdot)$,

*SingPoh.Toh@nottingham.edu.my; singpoh@gmail.com

would follow $v(L)v(M) = v(N)$. Hence, we may write down (1) based on the Mermin pentagram, i.e.,

$$\begin{aligned}
v(\sigma_x^1)v(\sigma_x^2)v(\sigma_z^3)v(\sigma_x^1\sigma_x^2\sigma_z^3) &= v(I), \\
v(\sigma_x^1)v(\sigma_z^2)v(\sigma_x^3)v(\sigma_x^1\sigma_z^2\sigma_x^3) &= v(I), \\
v(\sigma_z^1)v(\sigma_x^2)v(\sigma_x^3)v(\sigma_z^1\sigma_x^2\sigma_x^3) &= v(I), \\
v(\sigma_z^1)v(\sigma_z^2)v(\sigma_z^3)v(\sigma_z^1\sigma_z^2\sigma_z^3) &= v(I), \\
v(\sigma_z^1\sigma_z^2\sigma_z^3)v(\sigma_z^1\sigma_x^2\sigma_x^3)v(\sigma_x^1\sigma_z^2\sigma_x^3)v(\sigma_x^1\sigma_x^2\sigma_z^3) &= v(-I)
\end{aligned} \quad (1)$$

Note that $v(\cdot) = \pm 1$ for the ten operators and $v(\pm I) = \pm 1$. If the values of an observable revealed by measurements are pre-determined and noncontextual, as assumed by NCHV theories, then obviously (1) will lead to a contradiction. Every value function for a particular operator at the left of the equality signs occurs twice, thus the product of the values assigned at the left of equality signs will give 1, but the product of the values assigned at the right of the equality signs will give -1. The contradiction demonstrates that if the results of measurements are pre-determined, it must depend on the measurement contexts. In other words, the contradiction would not be exist if the pre-determined measurement values are allowed to be contextual. The KS theorem is thus proven.

In 1995, Kernaghan and Peres [16] generated five complete orthogonal basis in eight-dimensional real Hilbert space, $\Re^8$, based on the operators in Mermin pentagram. The five orthogonal octad generated are listed as $R_1$-$R_8$, $R_9$-$R_{16}$, $R_{17}$-$R_{24}$, $R_{25}$-$R_{32}$ and $R_{33}$-$R_{40}$, respectively in Table I.

TABLE I: Kernaghan and Peres' 40 rays. The symbol $\bar{1}$ denotes $-1$.

| | | | | | | | |
|---|---|---|---|---|---|---|---|
| $R_1$ | 10000000 | $R_{11}$ | $1\bar{1}1\bar{1}0000$ | $R_{21}$ | 00110011 | $R_{31}$ | $010\bar{1}010\bar{1}$ |
| $R_2$ | 01000000 | $R_{12}$ | $1\bar{1}\bar{1}10000$ | $R_{22}$ | $0011001\bar{1}\bar{1}$ | $R_{32}$ | $010\bar{1}0\bar{1}01$ |
| $R_3$ | 00100000 | $R_{13}$ | 00001111 | $R_{23}$ | $001\bar{1}001\bar{1}$ | $R_{33}$ | $100101\bar{1}0$ |
| $R_4$ | 00010000 | $R_{14}$ | $000011\bar{1}\bar{1}$ | $R_{24}$ | $001\bar{1}00\bar{1}1$ | $R_{34}$ | $100\bar{1}0110$ |
| $R_5$ | 00001000 | $R_{15}$ | $00001\bar{1}1\bar{1}$ | $R_{25}$ | 10101010 | $R_{35}$ | $10010\bar{1}10$ |
| $R_6$ | 00000100 | $R_{16}$ | $00001\bar{1}\bar{1}1$ | $R_{26}$ | $1010\bar{1}0\bar{1}0$ | $R_{36}$ | $100\bar{1}0\bar{1}\bar{1}0$ |
| $R_7$ | 00000010 | $R_{17}$ | 11001100 | $R_{27}$ | $10\bar{1}010\bar{1}0$ | $R_{37}$ | $0110\bar{1}001$ |
| $R_8$ | 00000001 | $R_{18}$ | $1100\bar{1}\bar{1}00$ | $R_{28}$ | $10\bar{1}0\bar{1}010$ | $R_{38}$ | $01\bar{1}01001$ |
| $R_9$ | 11110000 | $R_{19}$ | $1\bar{1}001\bar{1}00$ | $R_{29}$ | 01010101 | $R_{39}$ | $0\bar{1}101001$ |
| $R_{10}$ | $11\bar{1}\bar{1}0000$ | $R_{20}$ | $1\bar{1}00\bar{1}100$ | $R_{30}$ | $01010\bar{1}0\bar{1}$ | $R_{40}$ | $0\bar{1}\bar{1}0\bar{1}001$ |

By choosing 36 out of the 40 rays, Kernaghan and Peres [16] put forward a state independent proof of the KS theorem. Their proof are most economical in terms of the number of rays used. Only after almost seventeen years, the KS theorem with 40 rays are proposed in [17] which runs as follow. Let's consider the completeness relations, $\sum P_i = I$, in (2). Note that $P_i$ is projector obtained by taking the outer product of the corresponding $R_i$.

$$\begin{aligned}
P_1 + P_2 + P_3 + P_4 + P_5 + P_6 + P_7 + P_8 &= I, \\
P_9 + P_{10} + P_{11} + P_{12} + P_{13} + P_{14} + P_{15} + P_{16} &= I, \\
P_{17} + P_{18} + P_{19} + P_{20} + P_{21} + P_{22} + P_{23} + P_{24} &= I, \\
P_{25} + P_{26} + P_{27} + P_{28} + P_{29} + P_{30} + P_{31} + P_{32} &= I, \\
P_{33} + P_{34} + P_{35} + P_{36} + P_{37} + P_{38} + P_{39} + P_{40} &= I, \\
P_1 + P_2 + P_3 + P_4 + P_{13} + P_{14} + P_{15} + P_{16} &= I, \\
P_1 + P_2 + P_5 + P_6 + P_{21} + P_{22} + P_{23} + P_{24} &= I, \\
P_1 + P_3 + P_5 + P_7 + P_{29} + P_{30} + P_{31} + P_{32} &= I, \\
P_2 + P_3 + P_5 + P_8 + P_{33} + P_{34} + P_{35} + P_{36} &= I, \\
P_9 + P_{10} + P_{13} + P_{14} + P_{19} + P_{20} + P_{23} + P_{24} &= I, \\
P_9 + P_{11} + P_{13} + P_{15} + P_{27} + P_{28} + P_{31} + P_{32} &= I, \\
P_9 + P_{12} + P_{14} + P_{15} + P_{34} + P_{36} + P_{38} + P_{39} &= I, \\
P_{17} + P_{19} + P_{21} + P_{23} + P_{26} + P_{28} + P_{30} + P_{32} &= I, \\
P_{18} + P_{19} + P_{21} + P_{24} + P_{33} + P_{34} + P_{38} + P_{40} &= I, \\
P_{25} + P_{28} + P_{30} + P_{31} + P_{33} + P_{36} + P_{37} + P_{38} &= I. \quad (2)
\end{aligned}$$

It is obvious from (2) that there are 20 projectors that occur four times each, i.e., $P_1$, $P_2$, $P_3$, $P_5$, $P_9$, $P_{13}$, $P_{14}$, $P_{15}$, $P_{19}$, $P_{21}$, $P_{23}$, $P_{24}$, $P_{28}$, $P_{30}$, $P_{31}$, $P_{32}$, $P_{33}$, $P_{34}$, $P_{36}$ and $P_{38}$. The remaining 20 projectors occur twice each in (2). As QM also obeys sum rule, it means that if the mutually commuting operators $L$, $M$ and $N$ are related via $L + M = N$, the relationship $v(L) + v(M) = v(N)$ is also holds. Each of the completeness relation in (2) can thus be written in the form of $\sum v(P_i) = v(I)$, as shown in (3).

$$\begin{aligned}
v(P_1) + v(P_2) + \ldots + v(P_8) &= v(I), \\
&\vdots \\
v(P_{25}) + v(P_{28}) + \ldots + v(P_{38}) &= v(I). \quad (3)
\end{aligned}$$

As $v(P_i) = 0$ or 1, each of the projectors at the left of equality signs in (2) occurs either twice or four times, and one and only one projector in each completeness equation takes value 1, the sum of the values assigned at the left of equality signs in (3) will give an even number. On the other hand, because of $v(I) = 1$, the sum of the values assigned at the right of equality signs in (3) results in an odd number, which is 15 in this case. This is again a contradiction that refutes the possibility of the noncontextual pre-determined values for the observables of three-qubit quantum system.

All the projectors in (2) used for the proof are rank-1 projectors (rays in $\Re_8$). We now propose the KS theorem proof that uses rank-2 projectors (planes in $\Re_8$). Consider the completeness relation $\sum P_{i,j} = I$, where

$P_{i,j}$ is the rank-2 projectors. We convert (2) and get

$$P_{1,7} + P_{2,8} + P_{3,4} + P_{5,6} = I,$$
$$P_{9,12} + P_{13,16} + P_{14,10} + P_{15,11} = I,$$
$$P_{19,20} + P_{21,22} + P_{23,17} + P_{24,18} = I,$$
$$P_{28,27} + P_{30,29} + P_{31,25} + P_{32,26} = I,$$
$$P_{33,35} + P_{34,40} + P_{36,37} + P_{38,39} = I,$$
$$P_{1,2} + P_{3,4} + P_{13,16} + P_{14,15} = I,$$
$$P_{1,2} + P_{5,6} + P_{21,22} + P_{23,24} = I,$$
$$P_{3,5} + P_{1,7} + P_{30,29} + P_{31,32} = I,$$
$$P_{3,5} + P_{2,8} + P_{33,35} + P_{34,36} = I,$$
$$P_{14,10} + P_{9,13} + P_{19,20} + P_{23,24} = I,$$
$$P_{15,11} + P_{9,13} + P_{28,27} + P_{31,32} = I,$$
$$P_{9,12} + P_{14,15} + P_{38,39} + P_{34,36} = I,$$
$$P_{23,17} + P_{19,21} + P_{32,26} + P_{28,30} = I,$$
$$P_{24,18} + P_{19,21} + P_{34,40} + P_{33,38} = I,$$
$$P_{31,25} + P_{28,30} + P_{36,37} + P_{33,38} = I. \quad (4)$$

Note that now there are 30 rank-2 projectors that each occur twice in (4). Based on the sum rule again, the value functions for the rank-2 projectors also follow the same functional relations, thus we obtain

$$v(P_{1,7}) + v(P_{2,8}) + \ldots + v(P_{5,6}) = v(I),$$
$$\vdots$$
$$v(P_{31,25}) + v(P_{28,30}) + \ldots + v(P_{33,38}) = v(I). \quad (5)$$

On the one hand, $v(P_{i,j}) = 0, 1$ and each of the $P_{i,j}$ occurs twice in (4), it means that the sum of the values assigned to the projectors at the left of equality signs in (5) results in an even number. Again, bear in mind that one and only one projector in each completeness relation in (4) would be assigned value 1. On the other hand, as $v(I) = 1$, the sum of the values assigned to the 15 $v(I)$ at the right of equality signs in (5) is 15, an odd number. This is again a contradiction that can't be avoided if the pre-determined values of measurement are noncontextual. The refutation of NCHV theories is thus clearly demonstrated through our KS theorem proof. The KS theorem proofs with 36, 38 and 40 rank-1 projectors that generated by Kernaghan and Peres [16] were proposed before [16, 17]. Our state independent proof associated with the Mermin pentagram is the most economical in terms of the number of projectors required, i.e., it uses only 30 projectors. It is noteworthy that $\sum P_{i,j} = I$ in (4) is constructed from $\sum P_i = I$ in (2) and there are more than one way to couple a particular rank-1 projector with the other projectors in the same measurement context to obtain a rank-2 projector. There are 243 proofs with rank-2 projectors that constructed based on (2). Furthermore, the orthogonality relationships between the projectors in (4) can be depicted through hypergraph shown in Figure 2 if we relabel the rank-2 projectors in such a way that $1, \cdots, 20$ represent the first 20 rank-2 projectors in (4) and $21, \cdots, 30$ represent $P_{1,2}, P_{3,5}, P_{9,13}, P_{14,15}, P_{19,21}, P_{28,30}, P_{23,24}, P_{31,32}, P_{34,36}$ and $P_{33,38}$, respectively.

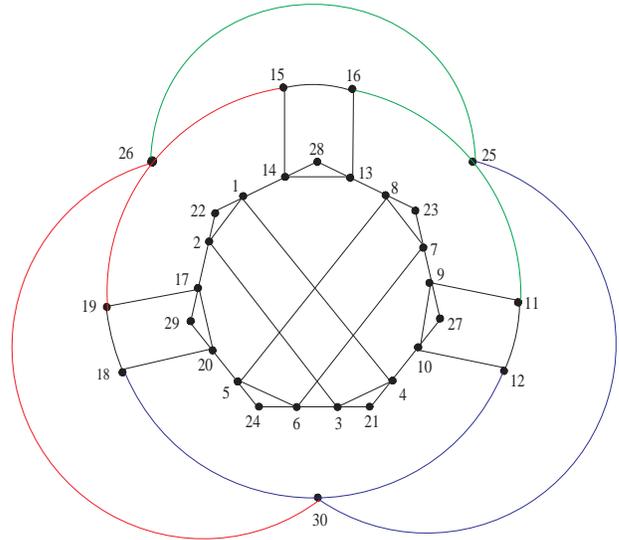

FIG. 2: (Color online) Hypergraph showing orthogonality relationships between 30 rank-2 projectors used to prove the KS theorem. The labeled vertices represent the projectors. Connected vertices are mutually orthogonal. Note that the red, green and blue curves connect four vertices in $\{15, 19, 26, 30\}$, $\{11, 16, 25, 26\}$ and $\{12, 18, 25, 30\}$, respectively.

In summary, we have proposed a state-independent proof of the KS theorem with 30 rank-2 projectors for a three-qubit system. We also introduced the KS theorem proof [13] with the 5 sets of compatible operators in Mermin pentagram and the proof with 40 rays [16] that generated from those 5 sets of operators in order to sketch the line of thought for getting our proof. The gist of all the proofs we presented, including ours, is that contextuality has to be introduced to establish consistency between pre-determined measurement results on quantum system and the predictions of quantum mechanics. Our proof can be cast in the form of testable inequality proposed by Cabello [3] and it requires the smallest number of projectors for eight-dimensional system associated with the Mermin pentagram.

This work is supported by the Ministry of Higher Education of Malaysia under the Fundamental Research Grant Scheme, FRGS/1/2011/ST/UNIM/03/1.